\documentstyle[preprint,eqsecnum,aps,prb]{revtex}
\tighten
\begin{document}
%
\preprint{EHU-FT-93/10}
\title{Parity Violation
in Aharonov-Bohm Systems: The Spontaneous Hall Effect}
\author{R.\ Emparan\cite{byline}}
\address{Departamento de F\'\i sica de la Materia Condensada\\
Universidad del Pa\'\i s Vasco, Apartado 644, 48080 Bilbao, Spain
}
\author{and\\
M.\ A.\ Valle Basagoiti}
\address{Departamento de F\'\i sica Te\'orica\\
Universidad del Pa\'\i s Vasco, Apartado 644, 48080 Bilbao, Spain
}
\maketitle
\begin{abstract}
We show how macroscopic manifestations of $P$ (and $T$)
symmetry breaking can arise in a simple system subject to
Aharonov-Bohm interactions. Specifically, we study the conductivity
of a gas of charged particles moving through a dilute array of flux
tubes. The interaction of the electrons with the flux tubes is taken
to be of a purely Aharonov-Bohm type. We find that the system
exhibits a non-zero transverse conductivity, i.e., a spontaneous
Hall effect. This is in contrast with the fact that the cross
sections for both scattering and  bremsstrahlung (soft photon
emission) of a single electron from a flux tube are invariant under
reflections. We argue that the asymmetry in the conductivity
coefficients arises from many-body effects. On the other hand, the
transverse conductivity has the same dependence on universal
constants that appears in the Quantum Hall Effect, a result that we
relate to the validity of the Mean Field approximation.
\end{abstract}

\narrowtext

\section{Introduction}

More than thirty years after it was first studied, the interaction
of flux tubes and charged particles --the Aharonov-Bohm (AB) problem
\cite{AB}-- continues to show a tantalizing richness. In recent
years it has received renewed attention due mainly to its
intimate relation with Chern-Simons (CS) field theory and the
ocurrence of fractional statistics in planar physics. One can
implement intermediate statistics on particles by attaching to them
appropriate `statistical' (i.e., fictitious) electric charge and
magnetic flux \cite{any}. Two such particles are then subject to AB
scattering, and acquire a statistical phase when moving around each
other. Varying the values of the charge and flux one can
continuously interpolate between bosons and fermions. Hence their
generic name, anyons.

On the other hand, CS field theory provides a very convenient way
to implement AB interactions between particles. The CS field
strength vanishes outside isolated singularities; thus, there are
no classical forces among particles. However, it is expected that a
mean field approximation, in which the particle is regarded as
moving in an average uniform magnetic field, should be valid
provided the flux per particle is small and that classical
trajectories enclose a large number of flux tubes. This approach
has been succesfully employed to calculate various properties of
the anyon gas \cite{LF,CWH}. In Ref.\,\onlinecite{CWH} a
self-consistency argument is given to support the mean field
approximation.

An important issue in fractional statistics and AB systems is that
of discrete (parity and time reversal) symmetry breaking. This
problem becomes of notorious relevance when one considers the
possible presence of anyonic excitations in high-$T_c$
superconductors, which are believed to present time reversal
non-invariance. Having these ideas in mind, March-Russell and
Wilczek \cite{MRW} sought for manifestations of $P$ and $T$
violation in CS theories. They found that the scattering cross
section of identical anyons (which is directly related to the AB
cross section) shows an asymmetry {\em provided an additional,
non-statistical interaction is included}. The parity violating term
arises from the interference between the amplitudes of AB
scattering and additional interaction. Caenepeel and MacKenzie
\cite{CMac} have used these results to test further the validity of
the mean field approximation. They consider the motion of one anyon
as the incoherent sum of two-particle scattering processes. They
find that, whenever there is an additional interaction switched on,
a typical trajectory is curved, its radius of curvature being large
for anyons near either bosons or fermions. In this regime, they
conclude, the mean field approximation can be justified. However,
the need for an additional interaction to show asymmetries remains
somewhat obscure. The results of Kiers and Weiss \cite{KW} suggest
that one can do without it and nevertheless find a handedness in
the scattering: particles are deflected by a regular lattice of
flux tubes if the full coherent scattering is taken into account.
As we shall see, these are not the only ways to make parity
breaking manifest.

As was already remarked in Ref.\
\onlinecite{MRW}, broken $T$ invariance should lead to asymmetries
in transport coefficients. A thorough investigation of the
experimentally observable consequences of $T$ and $P$ violation
in high-$T_c$ superconductors was carried out by Halperin,
March-Russell and Wilczek \cite{HMRW}. The standard Onsager
reciprocity relations do not apply when $P$ or $T$ are
broken, and anomalous transport phenomena are
expected. Working in the framework of an effective London
Lagrangian, they find that anyon systems should exhibit a Hall
conductivity in the absence of an applied magnetic field.

In this
paper we set out to study a simpler model which can be expected
to capture some essential aspects of $P$ (and $T$)
violation in macroscopic magnitudes. Specifically, the model
we analyze consists of a gas of charged particles moving through a
dilute random array of flux tubes. We shall work with a pure AB
interaction, not imposing {\em ab initio} an asymmetry in the
scattering cross-sections. We shall explicitly compute the
electrical (dc) conductivity of the system, expecting to find a
spontaneous Hall effect.  From the viewpoint of mean field theory
the graininess of the array should be unimportant. In this case one
would expect to find a transverse conductivity similar to that of
the Quantum Hall effect.

The outline of this paper is as follows: in Section II we set up
our model, in which flux tubes are regarded as impurities
distributed randomly, each acting independently on electrons.
After presenting the formalism for calculating
transport coefficients using field theory, we use it to
compute the longitudinal conductivity of the system. This
is essentially a calculation of vertex corrections
expressed in the form of a ladder equation. In Section III, the
parity violating, transverse conductivity is computed along these
lines. Although divergent terms appear in the calculations, all
physical quantities are finite. Section IV contains a discussion of
the results.

\section{Longitudinal conductivity through an array of flux tubes}

\subsection{The Model}

We start from a Lagrangian describing a free electron
gas interacting with an ensemble of flux tubes:
\begin{equation}\label{lagr}
{\cal L}={\cal L}_{\rm Fermi}-{\bf j}\cdot{\bf
A}-\rho\phi-\rho A^2/2m\,\,,
\end{equation}
where ${\cal L}_{\rm Fermi}$ is the free electron Lagrangian, ${\bf
j}$ and $\rho$ its corresponding current and density.

The flux
tubes are located at points ${\bf x}_a$, each being the source of a
vector potential ($\kappa$ is the flux per tube),
\begin{equation}\label{AFT} A_i({\bf x}-{\bf
x}_a)=\kappa{\epsilon_{ij}(x-x_a)_j\over |{\bf x}-{\bf x}_a|^2}
\end{equation}
and a scalar
potential,
\begin{equation}\label{FFT}
\phi({\bf x}-{\bf x}_a)
=g\delta^{(2)}({\bf x}- {\bf
x}_a)\,\,.
\end{equation}
Then ${\bf A}({\bf
x})=\sum_{{\bf x}_a}{\bf A}({\bf x}-{\bf x}_a)$, $\phi({\bf
x})=\sum_{{\bf
x}_a}\phi({\bf x}-{\bf x}_a)$. The flux tube density will be denoted $n_i$,
and the electron density $n_e$.

It is known that the presence of the contact interaction, Eq.\
(\ref{FFT}), is needed whenever one treats the AB problem within the
framework of CS theory \cite{BL,nos}. When its strength $g$ is
properly adjusted, the AB scattering amplitude is reproduced
\cite{BL}. We stress that this interaction has nothing to do with
the additional interaction introduced in Ref.\ \onlinecite{MRW}. In
our model, the right value for $g$ is
\begin{equation}\label{gk}
g=\pm \kappa/2m\,\,.
\end{equation}
This is {\em half} the value found in Refs.\ \onlinecite{BL,nos}.
The difference is due to the fact that we are considering the flux
tubes to be infinitely massive. Therefore the reduced mass of the
electron/flux tube system is twice that of a pair of identical
anyons. The choice of sign for $g$ determines the `hand' of the
interaction.

The vertex for interaction of flux tubes and electrons in
momentum space is depicted in Fig.\ 1. The corresponding amplitude
is
\begin{equation}\label{ABamp}
u({\bf p}+{\bf q},{\bf p})=g+i{\kappa\over m} {{\bf p} \times {\bf
q}\over q^2}\,.
\end{equation}
When $g$ is given by Eq.\ (\ref{gk}), this corresponds (apart from a
kinematical factor) to the AB scattering amplitude in
Born approximation. Flux tubes are assumed to be fixed objects with
no internal excitations, so the electrons scatter from them
elastically. This means that no frequency is carried by the
interaction lines.

As we shall show explicitly below, the `seagull' interaction
$\rho A^2/2m$ can be consistently neglected since it contributes
higher orders in perturbation theory.

We expect that this simple model
serves to approximate several
many-body systems where some kind of AB interaction is present. Of
course, more realistic interactions should be taken into account
before it is subject to experimental verification (for a detailed
study see Ref.\ \onlinecite{HMRW}). In this paper, it will be used
as a toy model for the study of the macroscopic effects that we
expect to arise reflecting the presence of the underlying AB
interaction. In particular, {\em can we make parity breaking show
up anyhow in transport coefficients?} Our answer to this question
will be on the positive.

\subsection{Computation of the Longitudinal Conductivity}

We shall regard flux tubes as impurities randomly distributed with
density $n_i<<1$. The first effect one expects is the appearance of
a finite longitudinal conductivity. This calculation will be done in
this section, following the treatment of Ref.\ \onlinecite{AGD}.
Having checked the validity of the method for this model, we shall
use it in the next section to find the transverse conductivity.

Consider then applying an external
electric field ${\bf E}^{\rm ext}=-\partial
{\bf A}^{\rm ext}/\partial t$ to the
system. A longitudinal conductivity will be induced, the
(linear) response being characterized by the conductivity tensor
\begin{equation}\label{cond}
J_i=\sigma_{ij}E^{\rm ext}_j=i\omega\sigma_{ij}A^{\rm ext}_j\,\,.
\end{equation}
On the
other hand, linear response theory yields the following relation
for the current induced in an electron gas by an applied vector
potential,
\begin{equation}\label{lrt}
J_i=-\Bigl({n_ee^2\over m}\delta_{ij}+\Pi_{ij}({\bf q},\omega)
\Bigr)A^{\rm ext}_j\,\,.
\end{equation}
{}From (\ref{cond}) we see that the dc conductivity can be
obtained from (\ref{lrt}) computing to $O(\omega)$ with ${\bf q}
=0$. $\Pi_{ij}({\bf q},\omega)$ is the (retarded) current-current
correlation function. It is best calculated in the imaginary time
formalism. This means that frequencies will be discrete
\begin{eqnarray}
\omega_n=&2\pi n/\beta&\qquad{\rm for\,\,bosons}\\
    =&2\pi (n+1)/\beta&\qquad{\rm for\,\,fermions}
\end{eqnarray}
while real frequency integrals $ (2\pi)^{-1}\int d\omega$ are
replaced by discrete sums $\beta^{-1}\sum_n$.
In imaginary time,
\begin{eqnarray}
\Pi_{ij}({\bf q},\omega_n)=&&\int_0^\beta d\tau\;e^{i\omega_n\tau}
\Pi_{ij}({\bf q},\tau)\,\,,\\
\Pi_{ij}({\bf q},\tau)=&&-\langle T\;j_i^\dagger ({\bf
q},\tau)j_j({\bf q},0)\rangle\,\,,
\end{eqnarray}
which can be calculated using standard diagrammatic techniques.
After performing the frequency sums one can analytically continue
to real retarded time by replacing
$i\omega_n\rightarrow\omega+i0^+$.

The free electron propagator (no impurities present) is
\begin{equation}\label{freeg}
G^0(\omega_n,{\bf p})=(i\omega_n-\xi_{\bf p})^{-1}\,\,,
\end{equation}
where $\xi_{{\bf p}}={\bf p}^2/2m-\epsilon_F$ is
the electron energy above the Fermi
level.

The effect of impurities is that the electron
propagator acquires an imaginary part corresponding to the
finite lifetime of excitations above the Fermi level. The Dyson
equation for the propagator leads to
\begin{equation}\label{g}
G(\omega_n,{\bf p})={1\over i\omega_n-
\xi_{\bf p}-i{\rm Im}\;\Sigma(\omega_n,{\bf p})}\,\,.
\end{equation}
The real part of the self-energy, ${\rm Re}\;
\Sigma(\omega_n,{\bf p})$ has been
absorbed in a renormalization of the Fermi level.

Since flux tubes are randomly distributed, we must average over the
position of each tube. After averaging, the Green's function in the
presence of impurities becomes translationally invariant: $\langle
G({\bf p},{\bf p}')\rangle=G(p)\delta^{(2)}({\bf p}-{\bf p}')$. On
the other hand, we shall be interested in excitations very near the
Fermi surface. This will help us simplify many calculations, since
all momenta will be strongly peaked around the Fermi value, $p_F$.

Our task now is to compute ${\rm Im}\;\Sigma$. Diagrams with only
one impurity line give (using translation invariance) a constant
which represents a shift of the Fermi energy, irrelevant for our
purposes. Now, under the assumptions that: a) the impurity density,
$n_i$, is low enough, b) scattering by impurities is weak (small
$\kappa$, Born approximation), then the main contribution to the
imaginary part of the self-energy is the one shown in Fig.\,2. For
low densities the interference between scattering from different
impurities is negligible, i.e., the scattering is incoherent.
Summing over the position of the impurities gives then a factor
$n_i$. On the other hand,  the seagull vertex, although of the same
order as the diagrams in Fig.\ 2, does not contribute to the
imaginary part.

These considerations lead to
\begin{eqnarray}
{\rm Im}\Sigma(\omega_n,{\bf p}) =&& n_i {\rm Im} \int{d^2k\over
(2\pi)^2}
 G^0(\omega_n,
{\bf k}) |u({\bf k},{\bf p})|^2
  \nonumber\\
  =&&n_im \int{d\xi_{\bf k}\over 2\pi} {\omega_n\over
\omega_n^2 +\xi_{\bf k}^2} \int{d\varphi\over 2\pi} |u(\varphi)|^2\,.
\end{eqnarray}
Here we have used
$|{\bf p}|,\;|{\bf k}|\sim p_F$ and assumed that, for ${\bf k}$ near
the Fermi
surface, $|u({\bf k},{\bf p})|^2$ is a slowly varying function. This
term is
\begin{equation}\label{uph}
|u({\bf k},{\bf p})|^2 = |u(\varphi)|^2 ={\kappa^2\over 2m^2}{1\over
1-\cos \varphi}\,\,.
\end{equation}
This is (up to a factor $m/2\pi v_F$) the differential cross section
$d\sigma/d\varphi$ for AB scattering \cite{AB} in the
Born approximation. It is even in the scattering angle.

Values of $\xi_{\bf k}$ far from the Fermi surface give negligible
contributions to the integral over energies. Therefore we can extend
the range of integration from $(-\epsilon_F,+\infty)$ to $(-\infty,
+\infty)$. One finds
\begin{equation}\label{ims}
{\rm Im}\;\Sigma(\omega_n,{\bf p})=-{{\rm sgn}\;\omega_n\over
2\tau}\,\,,
\end{equation}
where
\begin{equation}\label{tau}
\tau^{-1}=n_im\int{d\varphi\over 2\pi}|u(\varphi)|^2\,\,.
\end{equation}
The integrated cross section is $\sigma =(n_i v_F\tau)^{-1}$.
According to
this, $\tau$ is the mean time between collisions in the
Born approximation ($2\tau$ is the lifetime for an excitation near
the Fermi surface). Observe that $\tau^{-1}$ is divergent,
reflecting the long range of the AB interaction. The integral
in Eq.\ (\ref{tau}) has to be considered as regularized with a
cutoff for small $\varphi$, which eventually has to be
sent to zero. We do not write it explicitly since we shall find
that it disappears from physical macroscopic magnitudes.

Substitution of (\ref{ims}) into (\ref{g}) yields,
\begin{equation}\label{gdres}
G(\omega_n,{\bf p})={1\over i\omega_n-\xi_{\bf p}+i{\rm sgn}\;
\omega_n/2\tau} \,\,.
\end{equation}

The computation of the polarization tensor must take into
account the
fact that the average (over positions of flux tubes) of
two propagators does not equal the product of the separate averages.
The resulting terms can be conveniently included in the form of a
vertex term, as in Fig.\ 3. The vector vertex ${\bf\Gamma}$
takes account of diagrams where an impurity interacts with both the
upper and lower electron lines in Fig.\,3. Although it may seem at
first that these terms contain higher powers in the flux tube
concentration, $n_i$, this is not the case since they also contain
higher powers of ${\rm Im}\;\Sigma$. The seagull vertex can be
easily seen to yield a contribution of higher order in $\kappa$,
and therefore will be neglected.  Under the same assumptions made
above in the computation of ${\rm Im}\;\Sigma$, the main
contribution comes from the ladder diagrams (see Fig.\,4), and can
be found by solving the following integral equation:
\widetext
\begin{eqnarray}\label{ladder}
{\bf\Gamma}({\bf p},{\bf p}+{\bf q})=&& 2{\bf p} +{\bf q} +n_i\int
{d^2p'\over (2\pi)^2} \;G(\omega_p,{\bf p'})
 \biggl\{ |u({\bf p},{\bf p'})|^2 + ig {\kappa \over m}{{\bf q}
\times ({\bf p}-{\bf p'} )\over |{\bf p} -{\bf p'}|^2} \nonumber\\
&& +{\kappa^2 \over m^2}{({\bf p'} \times{\bf p} )\; {\bf q}\times
({\bf p} -{\bf p'} )
\over |{\bf p}-{\bf p'}|^4} \biggr\} {\bf\Gamma}({\bf p'},{\bf p'} +
{\bf q}) G
(\omega_p+\omega,{\bf p'} +{\bf q})\,\,.
\end{eqnarray}
(We are not writing the explicit frequency dependence of
${\bf\Gamma}$). In general, this equation is very difficult to solve.
However, we know that to
compute the longitudinal conductivity we must take ${\bf
q}=0$. The resulting equation is
\begin{equation}\label{ladder0}
{\bf\Gamma}^0({\bf p})= 2{\bf p}+n_i\int{d^2p' \over (2\pi)^2}p
\;G(\omega_p,{\bf
p'})|u({\bf p},{\bf p'})|^2 {\bf\Gamma}^0 ({\bf p'}) G(\omega_p+
\omega,{\bf p'})\,\,.
\end{equation}
Now, it is
clear that ${\bf\Gamma}^0 ({\bf p})\propto {\bf p}$. Write then
$
{\bf\Gamma}^0 ({\bf p})=(2+\Lambda){\bf p}\,\,,
$
where, for  $|{\bf p}|\sim p_F$, $\Lambda$ can be taken
as independent of $|{\bf p}|$.
Eq.\ (\ref{ladder0}) gives then the following equation for
$\Lambda$:
\begin{equation}\label{lamb1}
\Lambda=n_im(2+\Lambda)\int{d\xi_{\bf p'}\over 2\pi}G(\omega_p,
{\bf p'}) G(\omega_p+\omega,{\bf p'})  \int{d\varphi\over
2\pi}|u(\varphi)|^2{{\bf p}\cdot{\bf p'}\over p_F^2}
\,\,.
\end{equation}
\narrowtext
The integral over energies can be easily performed extending
the limits to $(-\infty,+\infty)$, and using contour integration. We
quote the result since it will be used repeatedly:
\begin{equation}\label{gg}
\int{d\xi_{\bf p'}\over 2\pi}G(\omega_p,{\bf p'})
G(\omega_p+\omega,{\bf p'})={\theta(-\omega_p) \theta
(\omega_p+\omega) \over \omega+1/\tau}\,,
\end{equation}
for $\omega >0$ (considering $\omega<0$ does not affect our final
results). Then
\begin{equation}\label{lamb2}
\Lambda={2\theta(-\omega_p)\theta (\omega_p+\omega)\over
\tau_1(\omega+1/\tau_{\rm tr})}\qquad(\omega>0)\,\,,
\end{equation}
with
\begin{eqnarray}
\tau^{-1}_1&=&n_im\int{d\varphi\over
2\pi}|u(\varphi)|^2\cos\varphi\,\,,
\\
\tau_{\rm tr}^{-1}&=&\tau^{-1}-\tau^{-1}_1={n_i\kappa^2 \over
2m}\,\,.
\end{eqnarray}
$\tau_{\rm tr}$ is the `transport time'
between collisions. The divergence in $\tau^{-1}_1$ is of the same
kind as the one in $\tau^{-1}$, so that their difference is
finite. Therefore, the dependence on the regularization of integrals
disappears from $\tau_{\rm tr}$.

Now we are ready to compute $\Pi_{ij}(\omega) \equiv \Pi_{ij} ({\bf
q} =0,\omega)$. From Fig.\,3(a) we read
\begin{eqnarray}\label{pij}
\Pi_{ij}(\omega)= &&{e^2\over
4m^2\beta}\sum_{\omega_p}\int{d^2p\over (2\pi)^2} \nonumber\\
&&\times 2p_i\;G(\omega_p,{\bf p})
\Gamma_j^0({\bf p})
G(\omega_p+\omega,{\bf p})\,\,.
\end{eqnarray}
Eq.\ (\ref{pij}) can now be solved.
Calculations are quite standard (see Ref.\,\onlinecite{AGD}). The
longitudinal conductivity is eventually found to be
\begin{equation}\label{slong}
\sigma_ L={n_ee^2\tau_{\rm tr}\over m}=2{n_ee^2 \over
n_i\kappa^2}\,\,.
\end{equation}
Remarkably, this is a non-zero, finite quantity, in spite of
the long range interaction. In contrast, if the interaction
with impurities were Coulombian, $\tau_{\rm tr}^{-1}$ would be
divergent, and therefore
$\sigma_ L \rightarrow 0$. However, the transverse nature of the AB
interaction makes
the divergence in $|u(\varphi)|^2$ milder.

\section{The transverse conductivity}

Here we shall use the techniques developed in the previous section
to make the presence of a transverse conductivity manifest.

A current perpendicular to the applied field must satisfy a relation of the
form
\begin{equation}\label{perpj}
J_i=\sigma_\perp\epsilon_{ij}E_j^{\rm ext}\,\,.
\end{equation}
Here, the presence of a transverse conductivity
$\sigma_\perp$ is a signal of $P$ and
$T$ spontaneous symmetry breaking. The relation $J_i=\sigma_L
E_i^{\rm ext}$ already breaks $T$ symmetry, but this is naturally
expected since dissipative effects are present. However, $T$
breaking is qualitatively different in (\ref{perpj}) because
the transverse current does not dissipate energy (${\bf
J}\cdot {\bf E}^{\rm ext} =0$).

We find it convenient to choose a gauge in which
the external field takes the form ${\bf E}^{\rm ext}=-\nabla\phi^
{\rm ext}$, or, in Fourier components, ${\bf E}^{\rm ext}=-i{\bf
q}\phi^{\rm ext}$. Therefore
\begin{equation}\label{jphi}
J_i=-i\sigma_\perp\epsilon_{ij}q_j\phi^{\rm ext}\,\,.
\end{equation}
This means that, since $J_i=-\Pi_{0i}\phi^{\rm ext}$, it
will suffice to compute the density-current correlation function
$\Pi_{0i}$ to $O(q)$, and then take
the limit $\omega\rightarrow 0$ (Fig.\,3(b)).

The vertex ${\bf\Gamma} ({\bf p}, {\bf p}+{\bf q})$ can be
expanded in  powers of ${\bf q}$ as
follows:
\begin{eqnarray}\label{ladexp}
{\bf\Gamma} ({\bf p}, {\bf p}+{\bf q}) =&&{\bf\Gamma}^0 ({\bf
p},{\bf p})+  {\bf\Gamma}^\epsilon ({\bf p}, {\bf p}+{\bf q})
\nonumber\\
&&+{\bf\Gamma}^{({\rm no}\;\epsilon)} ({\bf p}, {\bf p}+{\bf
q})+O(q^2)\,\,.
\end{eqnarray}
We have computed the zeroth order term ${\bf\Gamma}^0$ in
the previous  section.
The first order contribution has been split into a part
${\bf\Gamma}^\epsilon$  containing the vector ${\bf q}_\perp$
(defined to have components $\epsilon_{ij}q_j$), and another
${\bf\Gamma}^{({\rm no}\;\epsilon)}$, containing ${\bf q}$. We shall
be interested only in ${\bf \Gamma}^\epsilon$. The
integral equation for it can be obtained from Eq.\
(\ref{ladder}) (observe that the last term inside the brackets in
Eq.\ (\ref{ladder}) does not contribute to ${\bf \Gamma}^\epsilon$,
but to ${\bf\Gamma}^ {({\rm no}\;\epsilon)}$, since
$({\bf q}\times{\bf p}) ({\bf p'}\times{\bf p})= p^2{\bf q}\cdot
{\bf p'} - ({\bf q}\cdot{\bf p}) ({\bf p'}\cdot
{\bf p})$):
\widetext
\begin{eqnarray}\label{pvv}
{\bf\Gamma}^\epsilon({\bf p},{\bf p}+{\bf q})=&& n_i\int{d^2p'\over
(2\pi)^2} \;G(\omega_p,{\bf p'}) \biggl\{-ig{\kappa\over m}{ ({\bf
p}-{\bf p'})\cdot{\bf q}_\perp\over |{\bf p}-{\bf p'}|^2}
\;{\bf\Gamma}^0 ({\bf p'},  {\bf p'}) \nonumber\\
&&+|u({\bf p},{\bf p'})|^2 {\bf\Gamma}^\epsilon({\bf p'},{\bf
p'}+{\bf q})  \biggr\}
\;G(\omega_p+\omega,{\bf p'})\,\,.
\end{eqnarray}
\narrowtext
Changing the sign of $g$ (Eq.\ (\ref{gk})) would reverse the flow
of the transverse current.

The ansatz that allows us to solve (\ref{pvv}) is not very hard to
guess. After some examination, one is led to write
\begin{equation}\label{eans}
{\bf\Gamma}^\epsilon({\bf p},{\bf p} + {\bf q})={1\over p_F^2}{\bf
p}  ({\bf p}\cdot {\bf
q}_\perp) \Lambda_1+{\bf q}_\perp\Lambda_2\,\,,
\end{equation}
where, again, $\Lambda_1,\,\Lambda_2$ are independent of $|{\bf
p}|$.

The following integrals are needed to solve Eq.\,(\ref{pvv})
($\varphi$ is the
angle between ${\bf p}, {\bf p'}$):
\begin{eqnarray}
n_im\int&&{d\varphi\over
2\pi}|u(\varphi)|\;{\bf p'} ({\bf p'}\cdot{\bf q}_\perp)\nonumber\\
&&=({1\over \tau_1}-{1\over
\tau_{\rm tr}}){\bf p} ({\bf p}\cdot{\bf q}_\perp)
+{p_F^2 \over \tau_{\rm tr}} {\bf q}_\perp \,\,,
\end{eqnarray}
\begin{equation}
{n_i\kappa^2 \over m} \int{d\varphi\over
2\pi}\;{({\bf p}-{\bf p'})\cdot{\bf q}_\perp\over
|{\bf p}-{\bf p'}|^2} {\bf
p'} = {1\over \tau_{\rm tr}} \biggl({1\over p_F^2}{\bf p}({\bf
p}\cdot {\bf q}_\perp)
-{\bf q}_\perp\biggr)\,.
\end{equation}

Calculations are now straightforward, and lead (again for
$\omega>0$) to
\begin{eqnarray}\label{lsol}
\Lambda_1&=&-i{(\Lambda+2)\over 2}{\theta(-\omega_p) \theta
(\omega_p+  \omega)\over
\omega\tau_{\rm tr}+2}\,\,, \\
\Lambda_2&=&i{(\Lambda+2)\over
2}{\theta(-\omega_p)\theta (\omega_p+ \omega)\over \omega\tau_{\rm
tr}}\biggl(1-{1\over \omega\tau_{\rm tr}+2}\biggr)\,.
\end{eqnarray}

Now we can use the vertex ${\bf\Gamma}^\epsilon$ to find the part
of ${\bf \Pi}\equiv(\Pi_{0i})$ that yields parity violation. This is
(see Fig.\ 3(b)),
\begin{eqnarray}\label{tpol}
{\bf \Pi}^\epsilon &&={e^2\over 2m\beta} \sum_{\omega_p}
\int{d^2p\over (2\pi)^2}  \nonumber\\
&&\times G(\omega_p,{\bf p}) {\bf\Gamma}^\epsilon({\bf p},{\bf
p}+{\bf q}) G(\omega_p+\omega,{\bf p}+{\bf q})\,\,.
\end{eqnarray}
The ansatz (\ref{eans}) gives the following expression for ${\bf
\Pi}^\epsilon$ after integrating the angles:
\begin{equation}\label{finpol1}
{\bf \Pi}^\epsilon={\bf q}_\perp{e^2\over
2\beta}\sum_{\omega_p}({\Lambda_1 \over 2}+ \Lambda_2) \int
{d\xi_{\bf p}\over 2\pi} G(\omega_p,{\bf p}) G(\omega_p+\omega,{\bf
p})\,\,.
\end{equation}
Substituting the results above one finds,
\begin{equation}\label{finpol2}
{\bf \Pi}^\epsilon=i{e^2/8\pi\over 1-i\omega\tau_{\rm tr}} {\bf
q}_\perp\,\,.
\end{equation}
We have made the continuation $\omega \rightarrow -i\omega$. Highly
nontrivial cancellations of $\tau^{-1},\tau_1^{-1}$ have concurred
again to yield a finite result. Taking $\omega \rightarrow
0$ we find the transverse conductivity,
\begin{equation}\label{fins}
\sigma_\perp=e^2/8\pi\,\,,
\end{equation}
which is independent of $n_e,\,n_i,$ and $\kappa$. At first
sight, this seems to lead to the nonsensical result that taking
either $n_i$ or $\kappa$ to be zero, one still finds a finite
transverse conductivity. Of course, this is wrong: these limits
must be taken in Eq.\ (\ref{finpol2}) before $\omega \rightarrow
0$, and then one obtains $\sigma_\perp =0$. Also, one can check that
choosing $g=-\kappa/2m$, reverses the sign of the conductivity.

\section{Discussion}

There are two aspects of this result that deserve some
explanation: first, the appearance of the transverse
conductivity; second, its dependence on universal constants.

It may seem somewhat surprising to find a non-zero
transverse conductivity. Consider the motion of a single electron
in the presence of a flux tube. The differential scattering cross
section (\ref{uph}) is invariant under $\varphi \rightarrow
-\varphi$,  i.e., a typical electron trajectory is not deflected.
In order to make asymmetries appear, March-Russell and Wilczek
\cite{MRW} were forced to introduce an additional interaction (in
the form of a generic phase shift). Then parity is broken through
interference terms. This is explicitly illustrated by Suzuki {\em et
al.} in Ref.\ \onlinecite{SSBL}, where the additional interaction is
taken to be a hard-disk repulsion. The situation here is different,
since the interaction with the flux tubes is purely AB. Of course,
one could argue that we actually have another interaction, namely
that with the external electromagnetic field. But we have reasons to
believe that this by itself is not the origin of parity breaking. We
have analyzed a closely related problem: the bremsstrahlung for
emission of soft photons in the presence of a flux tube. Vertex
corrections have been taken into account. It turns out that, after
summing over the scattering angle of the emitted photon, the
corresponding cross section for electrons is invariant under parity.
It seems that an electron interacting with a flux tube emits
photons in a left-right symmetric way.

There is however another difference between our problem and that
of the scattering of a single electron from a single flux tube: it
is essential in our calculation to take into account the many-body
effects. Appart from yielding a damping of excited electrons,
many-body effects are present in the ladder resummation that we
perform to find the vertices. One can check that if only the first
term in the series is included, the transverse conductivity turns
out to be zero. Of course, it is not consistent to do so, as we have
argued before. We need to sum all the terms and it is
then that the asymmetry appears. Therefore parity breaking does
arise at the macroscopic level, although its microscopic origin
is hidden.

The universal dependence of the transverse conductivity is less
surprising once it is noticed that it is of the kind familiar from
the Quantum Hall effect. This seems to provide support to the mean
field approximation, since we find a behaviour analogous to that of
an electron gas in the presence of a magnetic field. However, we are
dealing with a perturbative approximation, so we do not expect to
find the Landau level structure responsible of the integer-spacing
in the Hall conductivity. Nor, evidently, can our model take into
account the null longitudinal resistivity that appears
simultaneously with it. Nevertheless, it is remarkable to find in
this simple model such a semi-quantitative agreement with mean field
theory.

The procedure we have described above can be modified to
compute other transport coefficients, such as thermal
conductivities; these will also show parity violation. However, it
seems more interesting to take one step further and consider the
conductivity of a gas of anyons with the full CS interaction taken
into account. Calculations with this model are much more difficult,
but the outcome would be a full description of transport phenomena
in anyonic systems.

\acknowledgements
We would like to thank C.\ Manuel and J.\ L.\ Ma\~ nes for helpful
comments. This work has been supported in part by CICYT Project No.
AEN93-0435 and by UPV 063.310-EB119-92. R.\ E.\ also acknowledges
the Ministerio de Educaci\'on y Ciencia (Spain) for an FPI grant.


\begin{figure}
\caption{Vertex for interaction between an
electron (continuous line) and a flux tube (square). The amplitude
is $u ({\bf p}
+{\bf q},{\bf p})=g+i{\kappa\over m} {{\bf p}\times {\bf q}\over
q^2}$.}
\end{figure}

\begin{figure}
\caption{The main contribution to the imaginary part of the self-energy.
The impurity lines give a contribution $|u({\bf k},{\bf p})|^2$.}
\end{figure}

\begin{figure}
\caption{(a) The current-current correlation function $\Pi_{ij}$. (b) The
parity violating density-current correlation function $\Pi_{0i}^\epsilon$.
The electron lines correspond to the propagator $G(\omega_n,{\bf p})$ dressed
by the interaction with impurities (fig.\ 2).} \end{figure}

\begin{figure}
\caption{Ladder equation for the vector vertex ${\bf\Gamma}$.}
\end{figure}

\end{document}